\documentstyle[aps,prl,multicol,epsf]{revtex}
\renewcommand{\narrowtext}{\begin{multicols}{2} \global\columnwidth20.5pc}
\renewcommand{\widetext}{\end{multicols} \global\columnwidth42.5pc}
\renewcommand{\v}[1]{{\bf #1}}

\newcommand{\s}{{\sigma}}

\newcommand{\gr}{{\nabla}}
\def\eqa{\begin{eqnarray}}
\def\eea{\end{eqnarray}}
\newcommand{\eq}{\begin{equation}}
\newcommand{\ee}{\end{equation}}
\newcommand{\nn}{\nonumber\\}
\newcommand{\Eq}[1]{Eq.~(\ref{#1})}
\newcommand{\p}{\partial}

\newcommand{\cC}{ {\cal C} }

\newcommand{\cH}{ {\cal H} }
\newcommand{\cM}{ {\cal M} }
\newcommand{\cL}{ {\cal L} }
\newcommand{\cP}{ {\cal P} }
\newcommand{\cS}{ {\cal S} }
\newcommand{\cCS}{ {\cal {CS}} }

\newcommand{\ra}{\rightarrow}

\begin{document}
\draft
\title{The Chern-Simons Invariant in the Berry Phase of a Two by Two Hamiltonian}
\author{Wu-Yi Hsiang}
\address{Department of Mathematics, Hong-Kong University of Science and \\Technology, Clear Water Bay,
Kowloon, Hong Kong.} \author{and}
\author{Dung-Hai Lee}
\address{Department of Physics, University of California, Berkeley, CA 94720.}
\maketitle
\begin{abstract}
The positive (negaive)-energy eigen vectors of the two by two Hamiltonian $H=\v r\cdot\vec{\s}$ where $\vec{\s}$
are the Pauli matrices and $\v r$ is a 3-vector, form a U(1) fiber bundle when $\v r$ sweeps over a manifold $\cM$
in the three dimensional parameter space of $\v r$ . For appropriately chosen base space $\cM$ the resulting fiber
bundle can have non-trivial topology. For example when $\cM=S^2\equiv\{\v r; |\v r|=1\}$ the corresponding bundle
has a non-zero Chern number, which is the indicator that it is topologically non-trivial. In this paper we
construct a two by two Hamiltonian whose eigen bundle shows a more subtle topological non-triviality over
$\cM=R^3\bigcup\{\infty\}$, the stereographic projection of $S^3$. This non-triviality is characterized by a
non-zero Chern-Simons invariant.
\end{abstract}
\vspace{0.2in}

\narrowtext Since it's discovery in 1984\cite{berry}, Berry's phase has played an important role in quantum
mechanics. For a simple example where Berry's phase occurs, consider the following two-level Hamiltonian \eq H(\v
r) =x\s_x+y\s_y+z\s_z\equiv\v r\cdot\vec{\sigma},\label{berry}\ee where $\s_{x,y,z}$ are the three components of
the Pauli matrices and $x,y,z$ are three real parameters. For a fixed $\v r=(x,y,z)$ the Hamiltonian given in
\Eq{berry} has two eigenvalues $E_{\pm}=\pm |\v r|$. Let us denote the corresponding normalized two-component
eigen vectors by $|\psi_{\pm}(\v r)>$.

For the rest of the paper let us focus on, say, the positive-energy eigen vector $|\psi_+(\v r)>$. Since the
transformation $|\psi_+(\v r)>\ra e^{i\theta}|\psi_+(\v r)>$ preserves the normalization and the eigen nature of
$|\psi_+(\v r)>$, there is a continuum of two-component vectors, each labeled by $\theta$, satisfying
$H|\psi_{+,\theta}(\v r)>=|\v r||\psi_{+,\theta}(\v r)>$.  For a given $\v r$ the internal space formed by all
$|\psi_{+,\theta}(\v r)>$ has the U(1) group structure. When $\v r$ varies over a manifold $\cM$ (henceforth
referred as the base space) the internal space sweeps out a ``fiber bundle''. Since the internal space has U(1)
symmetry this fiber bundle is called a U(1) bundle. In the following we shall refer to such fiber bundle as the
eigen bundle of \Eq{berry}.

A connection can be defined on the eigen bundle by first choosing a reference eigen vector $|\psi_+ (\v r)>$ at
each $\v r$ then defining the vector field \eq \v A_b(\v r)=\frac{1}{i}<\psi_+(\v r)|\gr\psi_+(\v r)>.\ee We note
that in order for $\v A_b$ to be well defined, the reference vector must vary continuously with $\v r$. The
connection $\v A_b$ defined above is not unique. Indeed, by performing the transformation $|\psi_+(\v r)>\ra
e^{i\theta(\v r)}|\psi_+(\v r)>$ we induce a ``gauge transformation'' on $\v A_b$: \eq\v A_b\ra\v
A_b+\gr\theta.\ee The Berry's phase caused by an adiabatic evolution of $\v r$ around a closed loop ${\cal{C}}$ is
given by \eq \gamma = \oint_{\cal{C}}d\v r\cdot\v A_b(\v r).\ee Obviously the Berry's phase is gauge invariant.

Let us first consider two-dimensional base spaces  $\cM$ that are closed surfaces. It turns out that if $\cM$
encloses the origin (for example $\cM=S^2=\{\v r;~|\v r|=1\}$), it is impossible to choose a gauge in which $\v
A_b(\v r)$ is non-singular everywhere. In order to obtain non-singular $\v A_b$ it is necessary to divide $\cM$
into a number of (overlapping) patches so that 1)$\v A_b$ is non-singular in each patch, and 2) in the overlapping
region of two patches the different $\v A_b$'s only differ by a gauge transformation. Historically a problem like
this was encountered when Dirac tried to write down the vector potential in the surrounding of a magnetic
monopole.\cite{dirac} It turns out that under the framework of quantum mechanics the condition 2) stated above
requires the strength of the monopole to be quantized.\cite{dirac,yang}

The fact that it is impossible to define an everywhere-nonsingular connection is the symptom of non-trivial
topology. In his seminal work Chern discovered a set of invariants to characterize such
non-triviality.\cite{chern} For the simple case we are considering the invariants reduce to a single number, the
Chern number: \eq \cC=\frac{1}{4\pi}\int_{\cM}d\v a\cdot\v B_b.\label{chn}\ee Here $\v B_b=\gr\times\v A_b$ is the
curvature associated with $\v A_b$.( Note that in order to have a well defined $\v B_b$ we need a {\it locally}
differentiable $\v A_b$.) For the eigen bundle of \Eq{berry} Berry has shown that\cite{berry} \eq \v B_b(\v
r)=\frac{1}{2}\frac{\hat{r}}{|\v r|^2}.\label{mp}\ee The right hand side of \Eq{mp} is the same as the magnetic
field produced by a monopole sitting at the origin. Thus if $\cM$ encloses the origin \Eq{chn} yields $\cC=1/2$
indicating the corresponding eigen bundle is non-trivial. On the other hand if $\cM $ does not enclose the origin,
then $\cC=0$ and the corresponding eigen bundle is trivial. The Chern number (\Eq{chn}) has a simple physical
interpretation - it is the total flux produced by the bundle curvature. In a proof similar to that given in
Ref.\cite{yang}, Chern showed that $\cC$ should be quantized to values $n/2$ where $n$ = integer.\cite{chern}

Since the Chern number can be larger than $1/2$, it is interesting to know what kind of Hamiltonian will exhibit
$\cC=n/2$ ($n>1$) eigen bundles.

One answer is given by the following $(n+1)\times (n+1)$ matrix  \eq H(\v r)=\v r\cdot\v S,\ee where $\v
S=(S_x,S_y,S_z)$ are the matrices representing the three generators of SU(2) in the higher spin ($S=n/2$)
representation. For example for $n=2$ we have \eqa &&S_x=\frac{1}{\sqrt{2}}\pmatrix{0&1&0\cr 1&0&1\cr
0&1&0},~S_y=\frac{1}{\sqrt{2}}\pmatrix{0&-i&0\cr i&0&-i\cr 0&i&0}\nn&&S_z=\pmatrix{1&0&0\cr 0&0&0\cr 0&0&-1}.\eea

There is another modification of \Eq{berry} which also leads to $\cC=n/2$ eigen bundle. Interestingly this time we
do not need to enlarge the dimension of the Hamiltonian matrix.  Consider the following $2\times 2$ matrix \eq
H(\v r)=\hat{h}(\v r)\cdot\vec{\s},\label{pon}\ee where $\hat{h}(\v r)$ is a suitably chosen unit vector field
that defines a mapping from $\cM$ (a closed two-dimensional surface) to $S^2$. It is known that such mappings can
be classified into homotopy classes each labeled by an integer \eq \cP=\int_{\Gamma}d\v a\cdot\v J.\ee Here $\v
J$, the dual of the Pontryagin form, is given by \eq J^{\mu}=\frac{1}{8\pi}\epsilon^{\mu\nu\lambda} \hat{h}\cdot
(\p_{\nu}\hat{h}\times\p_{\lambda}\hat{h}).\label{pot}\ee We will show later that by choosing a $\hat{h}(\v r)$
with $\cP=n$ in \Eq{pon} the corresponding eigen bundle has Chern number $\cC=n/2$.

The Chern number records the highest level of topological non-triviality. When the Chern number vanishes the eigen
bundle can still be non-trivial at a more subtle level. Let us consider a three-dimensional base space in which
$\cC=0$ for all closed surfaces, implying the absence of ``$\v B_b$-monopole''. In that case the ``flux lines'' of
$\v B_b$ form closed loops. There is a topological interesting situation in which these flux lines link with one
another. It is clear that this class of eigen bundles are topologically distinct from those whose curvature flux
loops do not link.

In 1974 Chern and Simons discovered an invariant, the Chern-Simons invariant, that quantifies this more subtle
topological non-triviality.\cite{cs} For a three-dimensional manifold $\cM$ the Chern-Simons invariant is given by
\eq \cCS=\frac{1}{4\pi}\int_{\cM} d^3r \v A_b\cdot\gr\times\v A_b.\label{csi}\ee In order for $\cCS$ to be gauge
invariant we require $\cM$ to be free of boundary. The topological information recorded by $\cCS$ is precisely the
linking between the $\v B_b$ flux lines. The fact that linking is only defined in three dimensions explains why
the Chern-Simons invariant requires a three dimensional base space.

 After realizing that there is another level of topological
non-triviality it is natural to ask whether one can modify \Eq{berry} so that the eigen bundle exhibits non-zero
$\cCS$. We shall prove that the Hamiltonian given by \Eq{pon} also works so long as $\hat{h}(\v r)$ is chosen
appropriately.

Now let us restrict ourselves to the case where the base space $\cM$ is $R^3\bigcup\{\infty\}$, the stereographic
projection of $S^3\equiv\{(x,y,z,w); x^2+y^2+z^2+w^2=1\}$ . In that case $\hat{h}(\v r)$ is a mapping from
$R^3\bigcup\{\infty\}$ to $S^2$. Due to the work of Hopf it is known that such mapping can also be classified into
homotopy classes. There also exists an integer, the Hopf invariant, that characterizes each class. The meaning of
a non-trivial Hopf map is revealed by the dual of its Pontryagin form (\Eq{pot}). For both the trivial and
non-trivial Hopf maps $\gr\cdot\v J=0$ everywhere hence the $\v J$ flux lines form closed loops. For a non-trivial
map the closed flux loops link with one another while for a trivial map they don't. Now we explain what is the
Hopf invariant. Since $\gr\cdot\v J=0$, there exists a vector field $\v A_h$ so that \eq \v
J=\frac{1}{2\pi}\gr\times\v A_h.\label{def}\ee The Hopf invariant is simply the Chern-Simons invariant for $\v
A_h$,\cite{wz} i.e., \eq \cH=\frac{1}{4\pi}\int_{\cM} d^3r \v A_h\cdot\gr\times\v A_h.\label{hf}\ee In Fig.1 we
plot $\v J$ in the central region of the following $\cH=1$ map\cite{shankar}: \eqa &&\v
r=r(\sin{\theta}\cos{\phi},\sin{\theta}\sin{\phi},\cos{\theta})\nn
&&\hat{h}=(\sin{\beta}\cos{\alpha},\sin{\beta}\sin{\alpha},\cos{\beta})\nn &&\alpha(\v r)=\Theta(r^2-1+i
2r\cos{\theta})-\phi\nn&&\beta(\v r)=2\Theta(\sqrt{(1-r^2)^2+4r^2\cos^2{\theta}}+i 2r|\sin{\theta}|).\label{map}
\eea In the above $\Theta(u+iv)$ is the angle of the U(1) phase factor $(u+iv)/|u+iv|$. The green and blue arrows
mark the x,y and z components of $\v J(\v r)$ respectively. The purpose of this plot is to illustrate that the
green arrows loop around the blue ones hence manifests the linking in of the $\v J$ lines.

In the rest of the paper we prove the following. \eqa 1. &&{\rm ~For~base~space~} \cM=S^2 {\rm
~the~eigen~bundle~of~}\Eq{pon}\nn &&{\rm has~} \cC=n/2 {\rm ~if~} \hat{h}(\v r) {\rm ~has~} \cP=n.\label{th1}\eea
\eqa &&2. {\rm ~ For~base~space~} \cM=R^3\bigcup\{\infty\} {\rm ~the~eigen~bundle~of~}\nn && \Eq{pon} {\rm ~has~}
\cCS=n {\rm ~if~} \hat{h}(\v r) {\rm ~has~} \cH=n.\label{th2}\eea

The proof rests on the identity that \eq \v B_b=2\pi\v J.\label{key}\ee which we now show. The Berry curvature is
given by \eq \v B_b=\frac{1}{i}\gr\times <\psi_{+}|\gr\psi_{+}>.\label{gph1}\ee Since $<\psi_{+}|\psi_{+}>=1$, it
follows that $<\gr\psi_{+}|\psi_{+}>+<\psi_{+}|\gr\psi_{+}>=0.$ Furthermore since
$<\gr\psi_{+}|\psi_{+}>=<\psi_{+}|\gr\psi_{+}>^*$ we conclude that $<\psi_{+}|\gr\psi_{+}>$ is pure imaginary.
Thus \eqa &&\v A_b=Im\left[<\psi_{+}|\gr\psi_{+}>\right]\nn &&\v B_{b}=Im\left[\gr\times
<\psi_{+}|\gr\psi_{+}>\right].\label{gph}\eea In component form the second of \Eq{gph} read \eqa
B^{\mu}&&=\epsilon^{\mu\nu\lambda}Im\left[< \p_{\nu}\psi_{+}|\p_{\lambda}\psi_{+}>\right]\nn
&&=\frac{\epsilon^{\mu\nu\lambda}}{2}Im\left[< \p_{\nu}\psi_{+}|\p_{\lambda}\psi_{+}>-<
\p_{\lambda}\psi_{+}|\p_{\nu} \psi_{+}>\right]\label{uy}.\eea To compute $<
\p_{\nu}\psi_{+}|\p_{\lambda}\psi_{+}>-< \p_{\lambda}\psi_{+}|\p_{\nu}\psi_{+}>$ we insert a complete set of
states ($I=\sum_{n=\pm}|\psi_n><\psi_n|$), and that gives \eqa &&<\p_{\nu}\psi_{+}|\p_{\lambda}\psi_{+}>-<
\p_{\lambda}\psi_{+}|\p_{\nu}\psi_{+}>\nn&&=\sum_{n=\pm}<\p_{\nu}\psi_{+}|\psi_n><\psi_n|\p_{\lambda}\psi_{+}>-[\nu\leftrightarrow
\lambda]\nn&&=<\p_{\nu}\psi_{+}|\psi_-><\psi_-|\p_{\lambda}\psi_{+}>-[\nu\leftrightarrow\lambda].\label{here}\eea
In reaching the last line we have used the fact that
$<\p_{\nu}\psi_{+}|\psi_+><\psi_+|\p_{\lambda}\psi_{+}>-[\nu\leftrightarrow\lambda]=0.$

To compute $<\psi_-|\p_j\psi_+>$ in \Eq{here} we express the eigen vector of $H(\v r')=H(\v r+\delta\v r)=H(\v
r)+\delta\v r\cdot\gr H$ in terms of those of $H(\v r)$ via first order perturbation theory. To the first order in
$|\delta\v r|$ we obtain \eqa |\psi_+(\v r')>&&=\left[|\psi_+>+\frac{<\psi_-|\delta\v r\cdot\gr
H|\psi_+>}{E_+-E_-}|\psi_->\right]\nn&&=\left[|\psi_+>+\frac{<\psi_-|\delta r_{\mu}
\p_{\mu}\hat{h}\cdot\vec{\sigma}|\psi_+>}{2}|\psi_->\right].\label{prime}\eea 
\Eq{prime} implies that \eqa
<\psi_-|\p_{\nu}\psi_+>&&=\frac{<\psi_-|\p_{\nu}\hat{h}\cdot\vec{\s}
|\psi_+>}{2}.\nn &&\eea As the result we have \eqa &&<\p_{\nu}\psi_{+}|\psi_-><\psi_-|\p_{\lambda}\psi_{+}>-[\nu
\leftrightarrow\lambda]\nn
&&=\frac{1}{4}[<\psi_+|\p_{\nu}\hat{h}\cdot\vec{\s}|\psi_-><\psi_-|\p_{\lambda}\hat{h}\cdot\vec{\s} |\psi_+>
-[\nu\leftrightarrow\lambda]\nn&&
=\frac{1}{4}\sum_{n=\pm}[<\psi_+|\p_{\nu}\hat{h}\cdot\vec{\s}|\psi_n><\psi_n|\p_{\lambda}\hat{h}\cdot\vec{\s}
|\psi_+> -[\nu\leftrightarrow\lambda]\nn&&
=\frac{1}{4}[<\psi_+|[\p_{\nu}\hat{h}\cdot\vec{\s},\p_{\lambda}\hat{h}\cdot\vec{\s}] |\psi_+>]\nn&&
=\frac{i}{2}\epsilon_{abc}(\p_{\nu}\hat{h}_a)(\p_{\lambda} \hat{h}_b)[<\psi_+|\s_c
|\psi_+>]\nn&&=\frac{i}{2}\epsilon_{abc}(\p_{\nu}\hat{h}_a)(\p_{\lambda} \hat{h}_b)\hat{h}_c= \frac{i}{2}
\hat{h}\cdot(\p_{\nu}\hat{h}\times\p_{\lambda} \hat{h}).\label{long}\eea In going from the second to the third
line of \Eq{long} we have used the fact that
$<\psi_+|\p_{\nu}\hat{h}\cdot\vec{\s}|\psi_+><\psi_+|\p_{\lambda}\hat{h}\cdot\vec{\s} |\psi_+>
-[\nu\leftrightarrow\lambda]=0$. Substituting \Eq{long} into \Eq{uy} we obtain \eqa
B^{\mu}&&=\epsilon^{\mu\nu\lambda}Im [<
\p_{\nu}\psi_{\pm}|\p_{\lambda}\psi_{\pm}>]=\frac{1}{4}\epsilon^{\mu\nu\lambda}
\hat{h}\cdot\p_{\nu}\hat{h}\times\p_{\lambda} \hat{h}\nn&&=2\pi J^{\mu}.\eea

After establishing \Eq{key} it is simple to prove (\ref{th1}) and (\ref{th2}). For (\ref{th1}) the Chern number is
given by \eq \cC=\frac{1}{4\pi}\int_{\cS}d\v a\cdot\v B_b=\frac{1}{2}\int_{\cS}d\v a\cdot\v J = \cP/2.\ee As the
result $\cP=n$ implies $\cC=n/2$. Now let us prove (\ref{th2}). \Eq{key} implies that \eq \gr\times \v A_b=2\pi
\frac{1}{2\pi}\gr\times \v A_h.\ee As the result $\v A_b$ and $\v A_h$ differ by a pure gauge at most \eq \v
A_b=\v A_h+\gr\phi.\ee Since \Eq{csi} is gauge invariant when $\cM=R^3\bigcup\{\infty\}$, we conclude \eqa
\cCS&&=\frac{1}{4\pi}\int_{S^3} d^3r \v A_b\cdot\gr\times\v A_b =\frac{1}{4\pi}\int_{S^3} d^3r \v
A_h\cdot\gr\times\v A_h \nn&&= \cH.\eea As the result $\cH=n$ implies $\cCS=n$.

In physics one often encounters Berry's phase when a system posses's ``fast'' and ``slow'' dynamic degrees of
freedom. When the fast degrees of freedom are ``integrated out'' it often produces, as part of the effective
action of the slow variables, a term (the Berry's phase) that is non-zero even when the slow variables change
adiabatically with time. Such term can fundamentally alter the behavior of the slow variables.

In the following we present an example where the fast degrees of freedom generate an effective action that is the
Hopf invariant of the slow variables. The model is a field theory in two space and one (Euclidean) time
dimensions. It consists of two fields: 1) a fermion field $\psi_{\s}(\v r,t)$, and 2) an unit vector field
$\hat{n}(\v r,t)$. The Lagrangian density is given as \eqa \cL &&=
\cL_{\psi}+\cL_{n}-g\hat{n}\cdot\bar{\psi}_{\alpha}\vec{\sigma}_{\alpha\beta}\psi_{\beta}\nn
\cL_{\psi}&&=\bar{\psi}_{\alpha}(\p_t-\mu)\psi_{\alpha} - \frac{1}{2m}\bar{\psi}_{\alpha}(\gr-i\v
A_{ex})^2\psi_{\alpha}\nn \cL_n&&=i~\Omega[\hat{n}]+\frac{K}{2}|\gr\hat{n}|^2.\eea In the above $m,g,c,\mu$ are
parameters of the model, $\v A_{ex}$ is the vector potential of an external magnetic field, i.e.,
$\p_xA_y-\p_yA_x=B$, and $\delta\Omega/\delta\hat{n}=\hat{n}\times\p_t\hat{n}$. Physically $\cL_{\psi}$ describes
fermions moving in an external magnetic field, and $\cL_n$ describes the dynamics of magnetic moments in a
ferromagnet. The last term in the first equation is the Zeeman coupling between the fermions and the magnetic
moments. By adjusting $\mu$ we can tune the density ($\bar{\rho}$) of the fermions so that \eq \bar{\rho}= k
\frac{B}{\phi_0},\label{int}\ee where $\phi_0=2\pi$ is the Dirac flux quantum and $k$ is an integer. When \Eq{int}
is satisfied, the ground state of the fermions forms a so-called  ``integer-quantum Hall liquid''.\cite{prange}
Let us further assume that $g$ is large so that locally the electron spins have to be in the direction of
$\hat{n}$. Under that condition integrating out the electrons produces a term $\frac{k}{4\pi}\int d^2xdt \v
A_h\cdot\gr\times\v A_h$, which is proportional to the Hopf invariant of the $\hat{n}(\v r,t)$. This term has the
effect of changing the spins and statistics of solitons (the skyrmions) in the $\hat{n}(\v r,t)$ field.\cite{wz}
\\

Acknowledgements: DHL is in debt to Geoffrey Lee for helping him to visualize the dual of the Pontryagin form of
the non-trivial Hopf map. We thank Qiang-Hua Wang for his help in  making Figure 1. DHL is supported by NSF grant
DMR 99-71503.

\noindent{\bf{Figure Caption}}
\\
\begin{figure}
\caption{The dual of the Pontryagin form in the central region of a $\cH=1$ map (\Eq{map}). The green and blue
arrows are the x,y and z components of $\v J$ respectively.}
\end{figure}

\widetext

\end{document}